\documentclass[aps,preprint,nofootinbib]{revtex4}
\usepackage{amsfonts}
\usepackage{amsmath}
\usepackage{amssymb}
\usepackage{relsize}
\usepackage{pdfpages}
\usepackage{float}
\def\correspondingauthor{\footnote{Corresponding author. E-mail address: Evgeny.Liverts@mail.huji.ac.il}}
\begin{document}
%\addcolour{Mauve}{0.5}{0.0}{0.5}
%\bibliographystyle{plainnat}
\title{Fock expansion for two-electron atoms. High order angular coefficients.}

\author{Evgeny Z. Liverts\correspondingauthor{}}
%\author{Evgeny Z. Liverts}
\affiliation{Racah Institute of Physics, The Hebrew University, Jerusalem 91904, Israel}

\author{Rajmund Krivec}
\affiliation{Department of Theoretical Physics, J. Stefan Institute, 1000 Ljubljana, Slovenia}
%\author{Nir Barnea}
%\affiliation{Racah Institute of Physics, The Hebrew University, Jerusalem 91904, Israel}

\begin{abstract}
The Fock expansion, which describes the properties of two-electron atoms near the nucleus, is studied.
The angular Fock coefficients $\psi_{k,p}(\alpha,\theta)$ with the maximum possible value of subscript $p$ are calculated on examples of the coefficients with $5\leq k \leq 10$. The presented technique makes it possible to calculate such angular coefficients for any arbitrarily large $k$.
The mentioned coefficients being leading in the logarithmic power series representing the Fock expansion may be indispensable for the development of simple methods for calculating the helium-like electronic structure.
The theoretical results obtained are verified by other suitable methods.
The Wolfram Mathematica is used extensively.
\end{abstract}

%\pacs{ }

\maketitle

\section{Introduction}\label{S0}

The properties of a two-electron atomic (helium-like) system with an infinitely massive nucleus of the charge $Z$ and non-relativistic energy $E$ are defined by the wave function (WF) $\Psi(r_1,r_2,r_{12})$, where
$r_1$ and $r_2$ are the electron-nucleus distances, and $r_{12}$ is the distance between the electrons.
The behavior of the ground state WF in
the vicinity of the nucleus located at the origin is determined by the Fock expansion \cite{FOCK}
\begin{equation}\label{1}
\bar{\Psi}(r_1,r_2,r_{12})\equiv\Psi(R,\alpha,\theta)=\sum_{k=0}^\infty R^k\sum_{p=0}^{[k/2]}\psi_{k,p}(\alpha,\theta)\ln^p R,
\end{equation}
where the hyperspherical coordinates $R,~\alpha$ and $\theta$ are defined by the relations:
\begin{equation}\label{2}
R=\sqrt{r_1^2+r_2^2},~~~~\alpha=2\arctan\left(\frac{r_2}{r_1}\right),~~~~\theta=\arccos\left(\frac{r_1^2+r_2^2-r_{12}^2}{2 r_1 r_2}\right).
\end{equation}
The convergence of expansion (\ref{1}) had been proven in Ref.\cite{MORG}.
The angular Fock coefficients (AFC) $\psi_{k,p}$
%(with $\psi_{0,0}=1$)
satisfy the Fock recurrence relation (FRR)
\begin{equation}\label{3}
\left[\Lambda^2-k(k+4)\right]\psi_{k,p}(\alpha,\theta)=h_{k,p}(\alpha,\theta)
\end{equation}
with the RHS of the form \cite{AB1}, \cite{LEZ4}:
\begin{equation}\label{4}
h_{k,p}=2(k+2)(p+1)\psi_{k,p+1}+(p+1)(p+2)\psi_{k,p+2}-2 V \psi_{k-1,p}+2 E \psi_{k-2,p}.
\end{equation}
The dimensionless Coulomb potential representing the electron-electron and electron-nucleus interactions is
\begin{equation}\label{5}
V\equiv \frac{R}{r_{12}}-Z\left(\frac{R}{r_1}+\frac{R}{r_2}\right)=
\frac{1}{\xi}
-\frac{2 Z\eta}{\sin \alpha},
\end{equation}
where we have introduced the important (in what follows) angular quantities:
\begin{equation}\label{6}
\xi=\sqrt{1-\sin \alpha \cos \theta},~~~\eta=\sqrt{1+\sin \alpha}.
\end{equation}
The hyperspherical angular momentum operator, projected on $S$ states, is defined as
\begin{equation}\label{7}
\Lambda^2=-4\left[\frac{\partial^2}{\partial\alpha^2}+2\cot\alpha\frac{\partial}{\partial\alpha}+ \frac{1}{\sin^2\alpha}\left( \frac{\partial^2}{\partial \theta^2}+\cot \theta \frac{\partial}{\partial\theta}\right)\right].
\end{equation}
It is clear that all circumnuclear features of the two-electron atoms (ions) are defined by the Fock expansion (\ref{1}).
There are a large number of methods for calculating the electronic structure of the two-electron atomic systems.
%\red{(it is needed to continue !)}.
An excellent review on this topic can be found in Refs. \cite{NAK,ROD,FOR,DRK,AB1}.
However, we are known only one technique that correctly represents the WF, $\Psi(r_1,r_2,r_{12})$ near the nucleus.
It is the so called correlation function hyperspherical harmonic  method (CFHHM) \cite{HM0,HM1,HM2}.
Unfortunately, the CFHHM is extremely difficult to implement.
The expansion in hyperspherical harmonics (HHs) provides the correct representation of the AFCs.
However, the HH-expansion is known to converge very slowly.
Although this method makes it possible to increase the convergence of the HH-expansion,
a sufficiently good accuracy requires a large HHs basis size, which, in turn, creates great computational difficulties.

Thus, the conclusion suggests itself that it would be extremely useful to develop a much simpler method for calculating the WF with correct behavior near the nucleus.
In this regard, we would like to emphasize the following important peculiarities of the Fock expansion (FE).
It follows from definition (\ref{1}) that the FE can be splitted into individual power series (lines) associated with definite power of $\textrm{ln} R$. In other words, the FE can be represented in the form:
\begin{eqnarray}\label{8}
\Psi=
(\ln R)^0\left(\psi_{0,0}+R \psi_{1,0}+R^2\psi_{2,0}+...\right)
~\nonumber~~~~~~~~~~~~~~~~~~~~~~~~~~~~~~~~~~~~~~~~~~~\\
+(\ln R)^1R^2\left(\psi_{2,1}+R \psi_{3,1}+R^2\psi_{4,1}+...\right)
~~~~~~\nonumber~~~~~~~~~~~~~~~~~~~~~~~~~~~~~\\
+(\ln R)^2R^4\left(\psi_{4,2}+R \psi_{5,2}+R^2\psi_{6,2}+...\right)
~\nonumber~~~~~~~~~~~~~~~~~~~~~~~~~~~\\
+(\ln R)^3R^6\left(\psi_{6,3}+R \psi_{7,3}+R^2\psi_{8,3}+...\right)
~\nonumber~~~~~~~~~~~~~~~~~~~~~\\
+(\ln R)^4R^8\left(\psi_{8,4}+R \psi_{9,4}+R^2\psi_{10,4}+...\right)
+...~~~~~~~~~
\end{eqnarray}
It is seen that the leading term of each line represents the product $(\ln R)^{k/2}R^k \psi_{k,k/2}(\alpha,\theta)$ with even $k$.
The first AFCs ($\psi_{0,0}=1$) corresponding to $k=0,2,4$ are well-known (see, e.g., \cite{AB1,LEZ4}):
\begin{equation}\label{9}
\psi_{1,0}= \frac{1}{2}\xi-Z \eta,
\end{equation}
\begin{equation}\label{10}
\psi_{2,1}= -\frac{Z(\pi-2)}{3\pi}(1-\xi^2),
\end{equation}
\begin{equation}\label{11}
\psi_{3,1} = \frac{Z(\pi-2)}{36\pi}\left[6Z \eta(1-\xi^2)+\xi (5\xi^2-6)\right],
\end{equation}
\begin{equation}\label{12}
\psi_{4,2} = \frac{Z^2(\pi-2)(5\pi-14)}{540\sqrt{\pi}}\left[Y_{40}(\alpha,\theta)+\sqrt{2}~Y_{42}(\alpha,\theta)\right].
\end{equation}
The normalized HHs are
\begin{equation}\label{13}
Y_{40}(\alpha,\theta)=\pi^{-3/2}(4\cos^2\alpha-1),~~~~~
Y_{42}(\alpha,\theta)=2\sqrt{2}\pi^{-3/2}\sin^2\alpha P_2(\cos \theta),~~~
\end{equation}
where $P_n(x)$ denotes the Legendre polynomials.

In this paper, we present the theoretical calculations of the AFCs $\psi_{5,2}(\alpha,\theta)$,  $\psi_{6,3}(\alpha,\theta)$, $\psi_{7,3}(\alpha,\theta)$, $\psi_{8,4}(\alpha,\theta)$ and $\psi_{9,4}(\alpha,\theta)$ included into the $k=4$, $k=6$ and $k=8$ "lines" of the expansion (\ref{8}), and also the AFC $\psi_{10,5}(\alpha,\theta)$ representing the leading term of $k=10$ "line".
%\red{BEGIN0}
It is important to note that all mentioned angular coefficients represent the AFCs $\psi_{k,p}$ with the maximum possible $p$ for a given $k$.
%\red{END0}

\section{Derivation of the angular Fock coefficient $\psi_{5,2}(\alpha,\theta)$}\label{S1}

The FRR (\ref{3})-(\ref{4}) for $k=5$ and $p=2$ reduces to the form
\begin{equation}\label{21}
\left(\Lambda^2-45\right)\psi_{5,2}(\alpha,\theta)=h_{5,2}(\alpha,\theta),
\end{equation}
where
\begin{equation}\label{22}
h_{5,2}(\alpha,\theta)=-2V\psi_{4,2}(\alpha,\theta).
\end{equation}
Using Eqs.(\ref{5})-(\ref{6}) and (\ref{12})-(\ref{13}), it is convenient to represent the RHS of Eq.(\ref{21}) in the form
\begin{equation}\label{23}
h_{5,2}(\alpha,\theta)=-\frac{Z^2(\pi-2)(5\pi-14)}{270\sqrt{\pi}}
\left[3\pi^{-3/2}(2h_1+h_2)-2Z(h_3+\sqrt{2}h_4)\right],
\end{equation}
where
\begin{equation}\label{24}
h_1=\frac{(1-\xi^2)^2}{\xi},~~~h_2=\frac{\cos(2\alpha)}{\xi},~~~h_3=\frac{\eta Y_{40}(\alpha,\theta)}{\sin \alpha}
~~~h_4=\frac{\eta Y_{42}(\alpha,\theta)}{\sin \alpha}.
\end{equation}
Accordingly, we obtain the solution of Eq.(\ref{21}) in the identical form
\begin{equation}\label{25}
\psi_{5,2}(\alpha,\theta)=-\frac{Z^2(\pi-2)(5\pi-14)}{270\sqrt{\pi}}
\left[3\pi^{-3/2}(2f_1+f_2)-2Z(f_3+\sqrt{2}f_4)\right],
\end{equation}
where the AFC-components $f_i$ satisfy the individual Fock recurrence relations (IFRR)
\begin{equation}\label{26}
\left(\Lambda^2-45\right)f_i=h_i.~~~~~~~~~~(i=1,2,3,4)
\end{equation}
We will sequentially find solutions to each of the IFRRs (\ref{26}) using
%, as a rule,
various methods presented in Ref.\cite{LEZ4}.

\subsection{Solution of the IFRR $(\Lambda^2-45)f_3=\eta Y_{40}/\sin\emph{} \alpha$ }\label{S1a}

Moving from simpler to more complex solutions, let's start with IFRR
\begin{equation}\label{27}
\left(\Lambda^2-45\right)f_3=h_3.
\end{equation}
The RHS $h_3\equiv h_3(\alpha)$ represents the function of only one angle variable $\alpha$. It was shown \cite{LEZ4} that the solution of the corresponding IFRR (\ref{27}) reduces to solution $g(\rho)=f_3(\alpha)$  of the inhomogeneous  differential equation
\begin{equation}\label{31}
(\rho^2+1)^2g''(\rho)+2\rho^{-1}(\rho^2+1)g'(\rho)+45g(\rho)=-\textrm{h}(\rho),
\end{equation}
where $\rho=\tan(\alpha/2)$, and
\begin{equation}\label{32}
\textrm{h}(\rho)\equiv h_3(\alpha)=\frac{(4\cos^2\alpha-1)\sqrt{1+\sin \alpha}}{\pi^{3/2}\sin \alpha}=
\frac{(\rho+1)(3\rho^4-10\rho^2+3)}{2\pi^{3/2}\rho(\rho^2+1)^{3/2}}.
\end{equation}
For convenience, we will solve the Eq.(\ref{31}) with the RHS $\textrm{h}(\rho)$ not containing the multiplier $\pi^{-3/2}$. The final solution $f_3$ will be multiplied by this factor.

Using the method of variation of parameters, one obtains \cite{LEZ4} the particular solution of Eq.(\ref{31}) in the form
\begin{equation}\label{33}
g^{(p)}(\rho)=v_{50}(\rho)\int\frac{u_{50}(\rho)\textrm{h}(\rho)d\rho}{(\rho^2+1)^2W_0(\rho)}-
u_{50}(\rho)\int\frac{v_{50}(\rho)\textrm{h}(\rho)d\rho}{(\rho^2+1)^2W_0(\rho)},
\end{equation}
where
\begin{equation}\label{34}
W_0(\rho)=-(\rho^2+1)/\rho^{2}.
\end{equation}
The independent solutions of the homogeneous equation associated with Eq.(\ref{31}) are \cite{LEZ4}:
\begin{equation}\label{35}
u_{50}(\rho)=\frac{(\rho^2+1)^{9/2}}{\rho}~_2F_1\left(4,\frac{7}{2};\frac{1}{2};-\rho^2\right)=
\frac{1-7\rho^2(3-5\rho^2+\rho^4)}{\rho(\rho^2+1)^{5/2}},
\end{equation}
\begin{equation}\label{36}
v_{50}(\rho)=(\rho^2+1)^{9/2}~_2F_1\left(4,\frac{9}{2};\frac{3}{2};-\rho^2\right)=
\frac{1-35\rho^2+21\rho^4-\rho^6}{7(\rho^2+1)^{5/2}},
\end{equation}
where $~_2F_1(...)$ is the Gauss hypergeometric function.
Substitution of Eqs.(\ref{34})-(\ref{36}) into the general representation (\ref{33}) yields
\begin{equation}\label{37}
g^{(p)}(\rho)=\frac{-7\rho\left\{\rho\left[5\rho(3\rho-4)(3\rho+5)-24\right]+23\right\}-23}{420\rho(\rho^2+1)^{5/2}}.
\end{equation}
The general solution of the inhomogeneous equation can be expressed as the sum of the general solution of the associated homogeneous (complementary) equation and the particular solution of the inhomogeneous equation, whence
\begin{equation}\label{37}
g(\rho)=g^{(p)}(\rho)+c_{u} u_{50}(\rho)+c_{v} v_{50}(\rho),
\end{equation}
where the coefficients $c_u$ and $c_v$ are currently undetermined.
To choose these coefficients, it is necessary to determine the behavior of all independent solutions on the boundaries of the domain $[0,\infty]$. We easily obtain:
\begin{equation}\label{38}
g^{(p)}(\rho)\underset{\rho\rightarrow 0}=-\frac{23}{420\rho}-\frac{23}{60}+\frac{451\rho}{840}+O(\rho^2),
\end{equation}
\begin{equation}\label{39}
g^{(p)}(\rho)\underset{\rho\rightarrow \infty}=-\frac{3}{4\rho}-\frac{1}{4\rho^2}+\frac{85}{24\rho^3}+O(\rho^{-4}),
\end{equation}
\begin{equation}\label{40}
u_{50}(\rho)\underset{\rho\rightarrow 0}=\frac{1}{\rho}-\frac{47\rho}{2}+O(\rho^3),~~~~~~~~~~~~~~~~~
\end{equation}
\begin{equation}\label{41}
u_{50}(\rho)\underset{\rho\rightarrow \infty}=-7+\frac{105}{2\rho^2}+O(\rho^{-4}),~~~~~~~~~~~~~
\end{equation}
\begin{equation}\label{42}
v_{50}(\rho)\underset{\rho\rightarrow 0}=1-\frac{15\rho^2}{2}+O(\rho^4),~~~~~~~~~~~~~~~~~
\end{equation}
\begin{equation}\label{43}
v_{50}(\rho)\underset{\rho\rightarrow \infty}=-\frac{\rho}{7}+\frac{47}{14\rho}+O(\rho^{-3}).~~~~~~~~~~~~~
\end{equation}
It is seen that the particular solution $g^{(p)}(\rho)$ is divergent at $\rho=0$, whereas the solutions of the homogeneous equation associated with Eq. (\ref{31}) are divergent,  at $\rho=0$ and $\rho=\infty$ for $u_{50}(\rho)$ and $v_{50}(\rho)$, respectively.
Thus, to avoid the divergence on the whole range of definition, one should set $c_u=23/420$ and $c_v=0$ in the general solution (\ref{37}).
Then the final physical solution becomes
\begin{eqnarray}\label{44}
f_3(\alpha)=-\frac{(\rho+1)(23\rho^4+22\rho^3-122\rho^2+22\rho+23)}{60\pi^{3/2}(\rho^2+1)^{5/2}}=
~\nonumber~~~~~~~~~~~~~~~~~~~~~~~~~~~~~~~\\
=-\frac{1}{60\pi^{3/2}}\left[11\sin \alpha+21 \cos(2\alpha)+2\right]\sqrt{1+\sin \alpha}.~~~~~~
\end{eqnarray}

\subsection{Solution of the IFRR $(\Lambda^2-45)f_4=\eta Y_{42}/\sin \alpha$ }\label{S1b}

It was shown in Ref.\cite{LEZ4} that the solution $f_4\equiv f_4(\alpha,\theta)$ of the IFRR
\begin{equation}\label{46}
\left(\Lambda^2-45\right)f_4=2\pi^{-3/2}\eta\sqrt{2} \sin\alpha P_2(\cos \theta)
\end{equation}
can be found in the form
\begin{equation}\label{47}
f_4=2\sqrt{2}~\pi^{-3/2} \sin^2\alpha P_2(\cos \theta)g_4(\rho),
\end{equation}
where the function $g_4(\rho)$ satisfies the equation
\begin{equation}\label{48}
(\rho^2+1)^2g_4''(\rho)+2\rho^{-1}[1+\rho^2+2(1-\rho^4)]g_4'(\rho)+13g_4(\rho)=-\textrm{h}_4(\rho),
\end{equation}
with
\begin{equation}\label{49}
\textrm{h}_4(\rho)=\frac{\eta}{\sin \alpha}=\frac{(\rho+1)\sqrt{\rho^2+1}}{2\rho}.
\end{equation}
Using the method of variation of parameters, one obtains \cite{LEZ4} the particular solution of Eq. (\ref{48}) in the form
\begin{equation}\label{50}
g_4^{(p)}(\rho)=v_{52}(\rho)\int\frac{u_{52}(\rho)\textrm{h}_4(\rho)d\rho}{(\rho^2+1)^2W_2(\rho)}-
u_{52}(\rho)\int\frac{v_{52}(\rho)\textrm{h}_4(\rho)d\rho}{(\rho^2+1)^2W_2(\rho)},
\end{equation}
where
\begin{equation}\label{51}
W_2(\rho)=-\frac{5}{\rho}\left(\frac{\rho^2+1}{\rho}\right)^5.
\end{equation}
The independent solutions of the homogeneous equation associated with Eq.(\ref{48}) are:
\begin{equation}\label{52}
u_{52}(\rho)=\frac{(\rho^2+1)^{13/2}}{\rho^5}~_2F_1\left(4,\frac{3}{2};-\frac{3}{2};-\rho^2\right)=
\frac{1+11\rho^2+99\rho^4-231\rho^6}{\rho^5\sqrt{\rho^2+1}},
\end{equation}
\begin{equation}\label{53}
v_{52}(\rho)=(\rho^2+1)^{13/2}~_2F_1\left(4,\frac{13}{2};\frac{7}{2};-\rho^2\right)=
\frac{231-99\rho^2-11\rho^4-\rho^6}{231\sqrt{\rho^2+1}}.
\end{equation}
Thus, the particular solution (\ref{50}) reduces to the form:
\begin{equation}\label{54}
g_4^{(p)}(\rho)=-\frac{11(21\rho^5+9\rho^4+\rho^2)+1}{2772\rho^5\sqrt{\rho^2+1}}.
\end{equation}
Considering the series expansions for $g_4^{(p)}(\rho)$, $u_{52}(\rho)$ and $v_{52}(\rho)$ on the boundaries of the range of definition ($\rho\in[0,\infty]$), it can be shown that function
\begin{equation}\label{55}
g_4^{(p)}(\rho)+\frac{1}{2772}u_{52}(\rho)=-\frac{\rho+1}{12\sqrt{\rho^2+1}}=-\frac{1}{12}\sqrt{1+\sin \alpha}
\end{equation}
represents the physical (finite) solution of Eq.(\ref{48}), and hence we finally obtain:
\begin{equation}\label{56}
f_4(\alpha,\theta)=-\frac{\sqrt{2}}{6\pi^{3/2}}(\sin\alpha)^2 \sqrt{1+\sin \alpha}~P_2(\cos \theta).
\end{equation}

\subsection{Solution of the IFRR $(\Lambda^2-45)f_1=(1-\xi^2)^2/\xi$}\label{S1c}

It is important to note that the RHS $h_1$ (see Eq.(\ref{24})) of the IFRR
\begin{equation}\label{57}
\left(\Lambda^2-45\right)f_1=h_1(\xi)
\end{equation}
is a function of $\xi$ (only) defined by Eq.(\ref{6}). For this case \cite{LEZ4, LEZ2} the solution of Eq.(\ref{57}) coincides with solution of the inhomogeneous differential equation
\begin{equation}\label{58}
(\xi^2-2)f_1''(\xi)+\xi^{-1}(5\xi^2-4)f_1'(\xi)-45f_1(\xi)=h_1(\xi).
\end{equation}
A particular solution of Eq. (\ref{58}) can be found by the method of variation of parameters in the form \cite{LEZ2}
\begin{equation}\label{59}
f_1^{(p)}(\xi)=\frac{1}{7\sqrt{2}}
\left[u_{5}(\xi)\int v_{5}(\xi)w(\xi)d\xi-v_{5}(\xi)\int u_{5}(\xi)w(\xi)d \xi\right],
\end{equation}
where
\begin{equation}\label{60}
w(\xi)=h_1(\xi)\xi^2\sqrt{2-\xi^2}.
\end{equation}
The linearly independent solutions of the homogeneous equation associated with Eq.(\ref{58}) are defined by the relations
\begin{equation}\label{61}
u_5(\xi)=\frac{P_{13/2}^{1/2}(\xi/\sqrt{2})}{\xi\sqrt[4]{2-\xi^2}}=\frac{2^{1/4}(8\xi^6-28\xi^4+28\xi^2-7)}{\sqrt{\pi(2-\xi^2)}},
\end{equation}
\begin{equation}\label{62}
v_5(\xi)=\frac{Q_{13/2}^{1/2}(\xi/\sqrt{2})}{\xi\sqrt[4]{2-\xi^2}}=
\frac{\sqrt{\pi}(-8\xi^6+20\xi^4-12\xi^2+1)}{2^{3/4}\xi}.
\end{equation}
where $P_\nu^\mu(x)$ and $Q_\nu^\mu(x)$ are the associated Legendre functions of the first and second kind, respectively.
Substitution of the representations (\ref{60})-(\ref{62}) into (\ref{59}) yields for the particular solution:
\begin{equation}\label{63}
f_1^{(p)}(\xi)=-\frac{1}{60}\xi(13\xi^4-30\xi^2+15).
\end{equation}
It can be verified that the particular solution $f_1^{(p)}(\xi)$ is finite on the whole range of definition ($\xi\in[0,\sqrt{2}]$), whereas the solutions of the homogeneous equation associated with Eq.(\ref{58}) are divergent at $\xi=\sqrt{2}$ ($\alpha=\pi/2, \theta=\pi$) and $\xi=0$ ($\alpha=\pi/2, \theta=0$) for $u_5(\xi)$ and $v_5(\xi)$, respectively.
Thus, we can conclude that the final physical solution of the IFFR (\ref{57}) coincides with the particular solution (\ref{63}), whence
\begin{equation}\label{64}
f_1=-\frac{1}{60}\xi(13\xi^4-30\xi^2+15)=-\frac{1}{60}\sqrt{1-\sin \alpha \cos \theta}
\left[\sin \alpha \cos \theta(4+13 \sin \alpha \cos \theta)-2\right].~~~~
\end{equation}

\subsection{Solution of the IFRR $(\Lambda^2-45)f_2=\cos(2\alpha)/\xi$} \label{S1d}

To solve the IFRR
\begin{equation}\label{65}
\left(\Lambda^2-45\right)f_2=h_2,
\end{equation}
with the RHS $h_2$ defined by Eq.(\ref{24}), first of all, it is necessary to recall the Sack's representation \cite{SACK} (see also \cite{AB1} and \cite{LEZ4}) for $\xi^{\nu}$ with $\nu=-1$:
\begin{equation}\label{66}
\xi^{-1}=\sum_{l=0}^\infty P_l(\cos \theta)\left(\frac{\sin \alpha}{2}\right)^l F_l(\rho),
\end{equation}
where
\begin{equation}\label{67}
F_l(\rho)=~_2F_1\left(\frac{l}{2}+\frac{1}{4},\frac{l}{2}+\frac{3}{4};l+\frac{3}{2};\frac{4\rho^2}{(\rho^2+1)^2}\right)=
\left\{ \begin{array}{c}~~\mathlarger{\mathfrak{F}_l(\rho) }~~~~~~~~~~~0\leq\rho\leq1 \\
\mathlarger{\mathfrak{F}_l(1/\rho) }~~~~~~~~~~~~\rho\geq 1
\end{array}\right.
\end{equation}
with
\begin{equation}\label{67a}
\mathfrak{F}_l(\rho)=
(\rho^2+1)^{l+\frac{1}{2}}.
\end{equation}
This enables us to present the RHS of Eq.(\ref{65}) in the form
\begin{equation}\label{68}
h_2\equiv\frac{\cos(2\alpha)}{\xi} =\sum_{l=0}^{\infty}P_l(\cos \theta)(\sin \alpha)^l \textrm{h}_l(\rho),
\end{equation}
where
\begin{equation}\label{69}
\textrm{h}_l(\rho)=2^{-l} F_l(\rho)\cos(2\alpha)=2^{-l} F_l(\rho)\left[1-\frac{8\rho^2}{(\rho^2+1)^2}\right].
\end{equation}
In turn, it was shown in Ref.\cite{LEZ4} that in case of the RHS is determined by Eq. (\ref{68}), the solution of the corresponding IFRR (\ref{65}) can be found in the form
\begin{equation}\label{70}
f_2(\alpha,\theta)=\sum_{l=0}^{\infty}P_l(\cos \theta)(\sin \alpha)^l \sigma_l(\rho),
\end{equation}
where the function $\sigma_l(\rho)$ satisfies the inhomogeneous differential equation
\begin{equation}\label{71}
(\rho^2+1)^2\sigma_l''(\rho)+2\rho^{-1}[1+\rho^2+l(1-\rho^4)]\sigma_l'(\rho)+(5-2l)(9+2l)\sigma_l(\rho)=-\textrm{h}_l(\rho).
\end{equation}
The linearly independent solutions of the homogeneous equation associated with Eq. (\ref{71}) are:
\begin{eqnarray}\label{72}
u_{5l}(\rho)=\frac{(\rho^2+1)^{l+9/2}}{\rho^{2l+1}}~_2F_1\left(4,\frac{7}{2}-l;\frac{1}{2}-l;-\rho^2\right)=
\frac{(\rho^2+1)^{l-5/2}}{\rho^{2l+1}}\times
~~~~~~~~~~~~~~~~~~~~~~~~~\nonumber~~~~~~~~\\
%\times
\left[\rho^6\left(1+\frac{120}{2l-5}-\frac{120}{2l-3}+\frac{24}{2l-1}\right)+
3\rho^4\left(1+\frac{40}{2l-3}-\frac{24}{2l-1}\right)+3\rho^2\left(1+\frac{8}{2l-1}\right)+1\right],~~~
\end{eqnarray}
\begin{eqnarray}\label{73}
v_{5l}(\rho)=(\rho^2+1)^{l+9/2}~_2F_1\left(4,\frac{9}{2}+l;\frac{3}{2}+l;-\rho^2\right)=
(\rho^2+1)^{l-5/2}\times
~~~~~~~~~~~~~~~~~~~~~~~~~\nonumber~~~~~~~~\\
%\times
\left[\rho^6\left(1-\frac{24}{2l+3}+\frac{120}{2l+5}-\frac{120}{2l+7}\right)+
3\rho^4\left(1+\frac{24}{2l+3}-\frac{40}{2l+5}\right)+3\rho^2\left(1-\frac{8}{2l+3}\right)+1\right].~~~~
\end{eqnarray}
The method of variation of parameters enables us to obtain the particular solution of the inhomogeneous differential equation (\ref{71}) in the form
\begin{equation}\label{74}
\sigma_l^{(p)}(\rho)=v_{5l}(\rho)\int\frac{u_{5l}(\rho)\textrm{h}_l(\rho)d\rho}{(\rho^2+1)^2W_l(\rho)}-
u_{5l}(\rho)\int\frac{v_{5l}(\rho)\textrm{h}_l(\rho)d\rho}{(\rho^2+1)^2W_l(\rho)},
\end{equation}
where
\begin{equation}\label{75}
W_l(\rho)=-\frac{2l+1}{\rho}\left(\frac{\rho^2+1}{\rho}\right)^{2l+1}.
\end{equation}
Note that due to different representations for the function $F_l(\rho)$ (see Eq.(\ref{67})) at values of $\rho$ less and greater than $1$, we obtain special representations for a particular solution in these two regions:
\begin{equation}\label{76}
\sigma_l^{(0)}(\rho)=\frac{(\rho^2+1)^{l-5/2}}{2^{l+1}(2l-1)(2l-3)}
\left[(2l-3)\rho^4-4(l-2)\rho^2-\frac{4l^2+4l-27}{3(2l-5)}\right],~~~~~0\leq\rho\leq 1~
\end{equation}
\begin{equation}\label{77}
\sigma_l^{(1)}(\rho)=\frac{\rho^{-2l-1}(\rho^2+1)^{l-5/2}}{2^{l+1}(2l+3)(2l+5)}
\left[\frac{4l^2+4l-27}{3(2l+7)}+4(l+3)\rho^2-(2l+5)\rho^4\right],~~~~~~~\rho\geq 1~
\end{equation}
It can be verified that both functions (\ref{76}) and (\ref{77}) have no singularities on their domains of definition.
On the other hand, function $u_{5l}(\rho)$ is singular at $\rho=0$, whereas $v_{5l}(\rho)$ is singular at $\rho=\infty$.
This means that one should search the general solution of Eq.(\ref{71}) in the form:
\begin{equation}\label{78}
\sigma_l(\rho)=\sigma_l^{(0)}(\rho)+c_{5l}^{(v)}v_{5l}(\rho),~~~~~~~~~~~~~~~~~~~0\leq\rho\leq 1~
\end{equation}
\begin{equation}\label{79}
\sigma_l(\rho)=\sigma_l^{(1)}(\rho)+c_{5l}^{(u)}u_{5l}(\rho),~~~~~~~~~~~~~~~~~~~~~~\rho\geq 1~
\end{equation}
Note that two coefficients $c_{5l}^{(v)}$ and  $c_{5l}^{(u)}$ are presently undetermined. To calculate them, we need to find two equations relating these coefficients.
The first equation is quite obvious. It follows from the condition that the representations (\ref{78}) and (\ref{79}) are coincident at the common point $\rho=1$, that is
\begin{equation}\label{80}
\sigma_l^{(0)}(1)+c_{5l}^{(v)}v_{5l}(1)=\sigma_l^{(1)}(1)+c_{5l}^{(u)}u_{5l}(1).
\end{equation}
This relationship reduces to the first desired equation:
\begin{equation}\label{81}
c_{5l}^{(u)}(2l+3)(2l+5)(2l+7)=c_{5l}^{(v)}(2l-5)(2l-3)(2l-1)+\frac{27-4l-4l^2}{3\times 2^{l+1}}.
\end{equation}
It can be verified that $\textrm{h}_l(0)=\textrm{h}_l(\infty)=2^{-l}$. It follows from this relations that
$\sigma_l^{(0)}(0)+c_{5l}^{(v)}v_{5l}(0)=\sigma_{l}^{(1)}(\infty)+c_{5l}^{(u)}u_{5l}(\infty)$.
It can be assumed that the last equation represents the second desired equation. However, this assumption turns out to be false, because it again leads to Eq.(\ref{81}).

We propose the following method to find the second desired equation.
Remind that any suitable function of the angles $\alpha$ and $\theta$ may be expanded into HHs since they form a complete set:
\begin{equation}\label{82}
f(\alpha,\theta)=\sum_{n=0(2)}^{\infty}\sum_{l=0}^{n/2}f_{n,l}Y_{n,l}(\alpha,\theta),
\end{equation}
where (see, e.g., \cite{AB1})
\begin{equation}\label{83}
f_{n,l}=\int f(\alpha,\theta)Y_{n,l}(\alpha,\theta)d\Omega
\end{equation}
with
\begin{equation}\label{84}
d\Omega=\pi^2 \sin^2\alpha\sin \theta d\alpha d\theta.~~~~~~~~~~~\alpha\in[0,\pi],~~\theta\in[0,\pi]
\end{equation}
For the function $f(\alpha,\theta)=f_2(\alpha,\theta)$ represented by Eq.(\ref{70}), the expansion coefficient with $n=2l$ becomes
\begin{equation}\label{85}
f_{2l,l}=\pi^2\int_0^\pi\int_0^\pi f_2(\alpha,\theta)Y_{2l,l}(\alpha,\theta)\sin^2\alpha \sin \theta d\alpha d\theta=
\frac{2\pi^2 N_{2l,l}}{2l+1}\left[K_0(l)+c_l^{(v)}K_v(l)+c_l^{(u)}K_u(l)\right],
\end{equation}
where
\begin{eqnarray}\label{86}
K_0(l)=\int_0^{\pi/2}(\sin \alpha)^{2l+2}\sigma_l^{(0)}(\rho)d\alpha+\int_{\pi/2}^{\pi}(\sin \alpha)^{2l+2}\sigma_l^{(\infty)}(\rho)d\alpha=
~~~~~~~~~~~~~~~~\nonumber~~~~~~~~~~~~\\
=\frac{\sqrt{2}\left(-24l^3-100l^2+198l +249\right)}{(2l-5)(2l-3)(2l+3)(2l+5)(2l+7)(2l+9)},~~~~~~~~
\end{eqnarray}
\begin{equation}\label{87}
K_v(l)=\int_0^{\pi/2}(\sin \alpha)^{2l+2}v_{5l}(\rho)d\alpha=\frac{2^{l+3/2}(2l-1)}{(2l+5)(2l+9)},~~~~~~~
\end{equation}
\begin{equation}\label{88}
K_u(l)=\int_{\pi/2}^{\pi}(\sin \alpha)^{2l+2}u_{5l}(\rho)d\alpha=\frac{2^{l+3/2}(2l+3)(2l+7)}{(2l-5)(2l-3)(2l+9)}.
\end{equation}
To derive the results (\ref{85})-(\ref{88}), we used the representation
$Y_{2l,l}(\alpha,\theta)=N_{2l,l}\sin^l\alpha P_l(\cos \theta)$
for the particular case of the HHs, and the orthogonality property for the Legendre polynomials.
It should be noted that the explicit form of the normalization constant $N_{2l,l}$ is not required.

On the other hand, expanding $f_2(\alpha,\theta)$ in HHs, and inserting this expansion into the LHS of the IFRR (\ref{65}), we obtain
\begin{equation}\label{89}
\left(\Lambda^2-45\right)f_2(\alpha,\theta)=\sum_{n=0(2)}^\infty\sum_{l=0}^{n/2}f_{n,l}\left[n(n+4)-45\right]Y_{n,l}(\alpha,\theta).
\end{equation}
To derive the last equation we used the fact that $Y_{n,l}(\alpha,\theta)$ is an eigenfunction of the operator $\Lambda^2$ with an eigenvalue equal to $n(n+4)$, that is
\begin{equation}\label{89a}
\Lambda^2 Y_{n,l}(\alpha,\theta)=n(n+4)Y_{n,l}(\alpha,\theta).
\end{equation}
The HH-expansion of the RHS of Eq. (\ref{65}) is
\begin{equation}\label{90}
h_2(\alpha,\theta)=\sum_{n=0(2)}^{\infty}\sum_{l=0}^{n/2}h_{n,l}Y_{n,l}(\alpha,\theta).
\end{equation}
Whence,
\begin{equation}\label{91}
f_{2l,l}=\frac{h_{2l,l}}{4l(l+2)-45}.
\end{equation}
Using again the Sack's representation (\ref{66})-(\ref{67}) and Eq.(\ref{83}), we obtain the expansion coefficient $h_{2l,l}$ in explicit form:
\begin{equation}\label{92}
h_{2l,l}=\pi^2\int_0^{\pi}\int_0^{\pi}h_2(\alpha,\theta)Y_{2l,l}(\alpha,\theta)\sin^2\alpha\sin \theta d\alpha d\theta
%~~~~~~~~~~~~~~~~\nonumber~~~~~~~~~~~~\\
=-\frac{2^{7/2}\pi^2 N_{2l,l} (4l^2+24l+19)}{(2l+1)(2l+3)(2l+5)(2l+7)}.~~~~
\end{equation}
Thus, inserting (\ref{92}) into the RHS of Eq. (\ref{91}) and equating the result to the RHS of Eq.(\ref{85}) we obtain the desired \emph{second equation} in the form:
\begin{equation}\label{93}
c_{5l}^{(u)}(2l+3)(2l+5)(2l+7)=-c_{5l}^{(v)}(2l-5)(2l-3)(2l-1)-2^{-l-1}(2l+1).
\end{equation}
Solving the system of two linear equations (\ref{81}) and (\ref{93}) gives the desired coefficients:
\begin{equation}\label{94}
c_{5l}^{(u)}=-\frac{2^{-l-1}(l+4)(2l-3)}{3(2l+3)(2l+5)(2l+7)},~~~~~
c_{5l}^{(v)}=\frac{2^{-l-1}(l-3)(2l+5)}{3(2l-5)(2l-3)(2l-1)}.
\end{equation}
%\red{BEGIN1}
It should be noted that the method described above for calculating the coefficients $c_{5l}^{(u)}$ and $c_{5l}^{(v)}$  is very reliable, but quite complex.
A much simpler method is based on the statement that the point $\rho=1$ represents the match point for the functions defined by Eqs. (\ref{78}) and (\ref{79}).
This means that not only these functions, but also their first (at least) derivatives must coincide at this point. Thus, the second required equation relating the coefficients $c_{5l}^{(u)}$ and $c_{5l}^{(v)}$ is:
\begin{equation}\label{94a}
\frac{d \sigma_l^{(0)}(\rho)}{d\rho}\Big{|}_{\rho=1}+c_{5l}^{(v)}\frac{d v_{5l}(\rho)}{d\rho}\Big{|}_{\rho=1}=
\frac{d \sigma_l^{(1)}(\rho)}{d\rho}\Big{|}_{\rho=1}+c_{5l}^{(u)}\frac{d u_{5l}(\rho)}{d\rho}\Big{|}_{\rho=1}.
\end{equation}
The solution of the system of two equations (\ref{81}) and (\ref{94a}) again gives the coefficients
% $c_{5l}^{(u)}$ and $c_{5l}^{(v)}$
defined by Eq.(\ref{94}).
%\red{END1}
Substituting these coefficients into the representations (\ref{78}) and (\ref{79}), we finally obtain:
\begin{equation}\label{95}
f_2(\alpha,\theta)=\frac{1}{6}\sum_{l=0}^\infty \frac{\zeta_l(\rho)P_l(\cos \theta)}{(2l-1)(2l+3)},
\end{equation}
where
\begin{equation}\label{96}
\zeta_l(\rho)=
\left\{ \begin{array}{c}\mathlarger{\chi_l(\rho) },~~~~~~~~~~0\leq\rho\leq1\\
\mathlarger{\chi_l(1/\rho)}, ~~~~~~~~~~~~~\rho\geq 1
\end{array}\right.
\end{equation}
with
\begin{equation}\label{96a}
\chi_l(\rho)=
\frac{\rho^l}{(\rho^2+1)^{5/2}}\left[\frac{(l-3)(2l-1)\rho^6}{2l+7}+9l \rho^4-9(l+1)\rho^2-\frac{(l+4)(2l+3)}{2l-5}\right].
\end{equation}
%Note that there is no $\sin^l\alpha $ in the representation (\ref{95}) because it is already included into the %function $\zeta_l(\rho)$ defined by Eqs.(\ref{96}) and (\ref{96a}).
%\red{BEGIN2}
It is clear that only the function (\ref{78}) is required for calculating the function $\chi_l(\rho)$.
Thus, in fact, we need to calculate only one coefficient $c_{5l}^{(v)}$ to define this function.
In this regard, it is important to emphasize that the representation (\ref{96}) reflects the fact that
the WF of a two-electron atomic system must preserve its parity with interchanging the electrons.
For the singlet S-states (which include the ground state) this means that the AFC and/or its component preserves its form (including the sign) under the transformation $\alpha\rightleftarrows \pi-\alpha$.
For the AFC-component $f_2(\alpha,\theta)$, represented by the series expansion (\ref{70}), this property corresponds (in terms of variable $\rho$) to the relationship:
\begin{equation}\label{96b}
\sigma_l^{(0)}(\rho^{-1})+c_{5l}^{(v)}v_{5l}(\rho^{-1})=\sigma_l^{(1)}(\rho)+c_{5l}^{(u)}u_{5l}(\rho).
\end{equation}
Elimination of the RHSs between equations (\ref{80}) and (\ref{96b}) for $\rho=1$ yields the identity,
whereas the use of Eq.(\ref{94a}) instead of Eq.(\ref{80}) yields the required equation:
\begin{equation}\label{96c}
\frac{d \sigma_l^{(0)}(\rho)}{d\rho}\Big{|}_{\rho=1}+c_{5l}^{(v)}\frac{d v_{5l}(\rho)}{d\rho}\Big{|}_{\rho=1}=
\frac{d \sigma_l^{(0)}(\rho^{-1})}{d\rho}\Big{|}_{\rho=1}+c_{5l}^{(v)}\frac{d v_{5l}(\rho^{-1})}{d\rho}\Big{|}_{\rho=1}.
\end{equation}
Solution of the last equation gives the coefficients $c_{5l}^{(v)}$ presented by Eq.(\ref{94}).
Note that the coefficient $c_{5l}^{(u)}$ can then be calculated by the use of Eq.(\ref{80}) if needed.
%\red{END2}

In the general case, we cannot sum the infinite series (\ref{95}), to obtain the function $f_2(\alpha,\theta)$ in an explicit closed form. However, this can be done for some special angles $\alpha$ and/or $\theta$.
For example, it is worth noting that the angles $\theta=0,\pi$ correspond to the collinear configuration \cite{LEZ6} of the two-electron atomic system in question. For these cases we obtain
\begin{equation}\label{97}
f_2(\alpha,0)=\pm\frac{(\rho-1)(12\rho^4-13\rho^3-88\rho^2-13\rho+12)}{90(\rho^2+1)^{5/2}},
\end{equation}
\begin{equation}\label{98}
f_2(\alpha,\pi)=-\frac{(\rho+1)(12\rho^4+13\rho^3-88\rho^2+13\rho+12)}{90(\rho^2+1)^{5/2}}.
\end{equation}
Sign "$+$" in Eq.(\ref{97}) corresponds to $0\leq\alpha\leq\pi/2$ ($0\leq \rho\leq 1$), whereas "$-$" to $\pi/2\leq\alpha\leq\pi$ ($ \rho\geq 1$).
The list of special $\theta$-angles can be supplemented with an intermediate angle $\theta=\pi/2$:
\begin{equation}\label{99}
f_2\left(\alpha,\frac{\pi}{2}\right)=-\frac{2(\rho^4-3\rho^2+1)}{15(\rho^2+1)^2}.
\end{equation}
It is worth noting that for the important cases of the nucleus-electron and electron-electron coalescence, Eq.(\ref{95}) respectively reduce to:
\begin{equation}\label{100}
f_2\left(0,\theta\right)=-\frac{2}{15},~~~~~~~~~~~~~f_2\left(\frac{\pi}{2},0\right)=0.
\end{equation}
To derive the results (\ref{97})-(\ref{100}) we used the relationships:
\begin{equation}\label{101}
P_n(0)=\sqrt{\pi}\Gamma^{-1}\left(\frac{1-n}{2}\right)\Gamma^{-1}\left(\frac{n}{2}+1\right),~~~~~
P_n(1)=1,~~~~~P_n(-1)=(-1)^n,
\end{equation}
where $\Gamma(x)$ is the gamma function.
%~\red{BBBBBBBBBBBBBBBBBBBBBBBBBBBBBBB}

\section{Derivation of the angular Fock coefficient $\psi_{6,3}(\alpha,\theta)$}\label{S2}

We start this section by considering the FRR (\ref{3})-(\ref{4}) for $k=6$ and $p=2$:
\begin{equation}\label{102}
\left(\Lambda^2-60\right)\psi_{6,2}=48 \psi_{6,3}-2 V \psi_{5,2}+2E \psi_{4,2}.
\end{equation}
Next, let's expand each function in Eq.(\ref{102}) into HHs, using Eq.(\ref{82}). This gives
\begin{equation}\label{103}
\psi_{k,p}=\sum_{n=0(2)}^\infty \sum_{l=0}^{n/2}c_{nl}^{(k p)}Y_{n,l}(\alpha,\theta),
\end{equation}
with $\{k,p\}=\{6,3\},\{6,2\},\{4,2\}$, and
\begin{equation}\label{104}
V \psi_{5,2}=\sum_{n=0(2)}^\infty \sum_{l=0}^{n/2}\textrm{f}_{nl}Y_{n,l}(\alpha,\theta),
\end{equation}
where the dimensionless potential $V$ is defined by Eq.(\ref{5}), whereas the expansion coefficient $\textrm{f}_{nl}$ can be calculated by the formula
\begin{equation}\label{105}
\textrm{f}_{nl}=\int V \psi_{5,2} Y_{n,l}(\alpha,\theta)d\Omega,
\end{equation}
according to Eq.(\ref{83}).

It follows from Eq.(\ref{4}) that $h_{k,k/2}=0$ for even $k$.
Using additionally Eq.(\ref{89a}), we can conclude that the AFC $\psi_{k,k/2}$ (with even $k$) represents the linear combination of the HHs, $Y_{k,l}(\alpha,\theta)$. Whence,
\begin{equation}\label{106}
c_{nl}^{(63)}=0~~~~ for~~~~n\neq 6,
\end{equation}
\begin{equation}\label{107}
c_{nl}^{(42)}=0~~~~ for~~~~n\neq 4.
\end{equation}
Equating the coefficients for the HHs, $Y_{6,l}(\alpha,\theta)$ in both sides of Eq.(\ref{102}), we obtain:
\begin{equation}\label{108}
0=48 c_{6l}^{(63)}-2\textrm{f}_{6l}.
\end{equation}
Whence (using additionally Eq.(\ref{105})),
\begin{equation}\label{109}
c_{6l}^{(63)}=\frac{1}{24}\int V \psi_{5,2}Y_{6,l}(\alpha,\theta)d \Omega.
\end{equation}
Note that the LHS of Eq.(\ref{108}) equals zero, because $(\Delta^2-60)Y_{6,l}=0$ as follows from Eq.(\ref{89a}).
%\red{(Conclusions regarding the choice between $Y_{6,l}$  according to electrons permutation!!)}

Thus, according to Eqs.(\ref{103}) and (\ref{106}) the AFC $\psi_{6,3}(\alpha,\theta)$ represents a linear combination of four HHs, $Y_{6,l}(\alpha,\theta)$ with $l=0,1,2,3$. The contribution of each HH is determined by the coefficient $c_{6l}^{(63)}$ given by Eq.(\ref{109}). 
%\red{BEGIN3} 
However, it is easy to prove that only the coefficients with odd values of $l$ are nonzero for $\psi_{6,3}(\alpha,\theta)$.
Indeed, this has already been mentioned in Sec. \ref{S1d} that the WF of a two-electron atom/ion must preserve its parity with interchanging the electrons. For the singlet S-states this means that only the HHs, which preserve the sign under transformation $\alpha \leftrightarrows \pi-\alpha$, differ from zero in the expansion of the WF, and hence in the expansion of any AFC. 
%\red{END3}
In turn, it is easy to show that only $Y_{n,l}(\alpha,\theta)$ with even values of $(n/2-l)$ satisfy the above property.
Whence, the AFC in question, becomes
\begin{equation}\label{110}
\psi_{6,3}(\alpha,\theta)=a_{61}Y_{6,1}(\alpha,\theta)+a_{63}Y_{6,3}(\alpha,\theta),
\end{equation}
where we denoted $a_{6l}\equiv c_{6l}^{(63)}$ ($l=1,3$) for convenience and simplicity, and where the normalized HHs are
\begin{equation}\label{110a}
Y_{6,1}(\alpha,\theta)=\frac{2\left[\sin \alpha+3 \sin(3\alpha)\right]\cos \theta}{\pi^{3/2}\sqrt{5}}, ~~~~~
Y_{6,3}(\alpha,\theta)=\frac{8\sin^3\alpha P_3(\cos \theta)}{\pi^{3/2}\sqrt{5}}.~~~~~
\end{equation}
Using formula (\ref{109}) and taking into account the representations (\ref{5}) and (\ref{25}) for the dimensionless potential $V$ and the AFC $\psi_{5,2}$, respectively, we can represent the desired coefficients in the form
\begin{equation}\label{111}
a_{6l}=-\frac{(\pi-2)(5\pi-14)}{6480}\left(I_{l,4}Z^4+I_{l,3}Z^3+I_{l,2}Z^2\right),~~~(l=1,3)
\end{equation}
where
\begin{equation}\label{112}
I_{l,4}=4\pi^{3/2}\int_0^\pi \int_0^\pi\left[f_3(\alpha,\theta)+\sqrt{2}f_4(\alpha,\theta)\right]\eta Y_{6,l}(\alpha,\theta)
\sin \alpha \sin \theta d\alpha d\theta,
\end{equation}
\begin{eqnarray}\label{113}
I_{l,3}=-2\int_0^\pi \int_0^\pi\left\{\frac{3\eta\left[2f_1(\alpha,\theta)+f_2(\alpha,\theta)\right]}{\sin \alpha}
+\frac{\pi^{3/2}\left[f_3(\alpha,\theta)+\sqrt{2}f_4(\alpha,\theta)\right]}{\xi}\right\}\times
~~~\nonumber~~\\
\times Y_{6,l}(\alpha,\theta)\sin^2 \alpha \sin \theta d\alpha d\theta,
\end{eqnarray}
\begin{equation}\label{114}
I_{l,2}=3\int_0^\pi \int_0^\pi\left[2f_1(\alpha,\theta)+f_2(\alpha,\theta)\right]\xi^{-1}Y_{6,l}(\alpha,\theta)
\sin^2 \alpha \sin \theta d\alpha d\theta.
\end{equation}
%It is easy to show that the integral $I_{l,4}$ vanishes both for $l=1$ and $l=3$.
To calculate the integral (\ref{113}), it is useful to separate the contributions which include the functions $f_1,f_3,f_4$ represented by the explicit closed expressions, and the function $f_2$ represented by the infinite series (\ref{95}). We obtain:
\begin{equation}\label{118}
I_{l,3}=I_{l,3}^{(134)}-6 I_{l,3}^{(2)},
\end{equation}
where
\begin{eqnarray}\label{119}
I_{l,3}^{(134)}=-2\int_0^\pi \int_0^\pi\left\{\frac{6\eta f_1(\alpha,\theta)}{\sin \alpha}
+\frac{\pi^{3/2}\left[f_3(\alpha,\theta)+\sqrt{2}f_4(\alpha,\theta)\right]}{\xi}\right\}
Y_{6,l}(\alpha,\theta)\sin^2 \alpha \sin \theta d\alpha d\theta,
~\nonumber~\\
\end{eqnarray}
\begin{equation}\label{120}
I_{l,3}^{(2)}=\int_0^\pi \int_0^\pi f_2(\alpha,\theta)\eta
Y_{6,l}(\alpha,\theta)\sin \alpha \sin \theta d\alpha d\theta.
\end{equation}
The integrals (\ref{119}) can be taken in explicit (closed) form that gives:
\begin{equation}\label{121}
I_{1,3}^{(134)}=\frac{3(45\pi-122)}{35 \pi^{3/2}\sqrt{5}},~~~~~~~~~I_{3,3}^{(134)}=\frac{245\pi-816}{70 \pi^{3/2}\sqrt{5}}.
\end{equation}
%It should be noted that a closed (explicit) form is not currently determined only for the integrals $I_{l,3}^{(2)}$.
The problem of calculating the  integrals (\ref{120}) is that the corresponding integrands contain the function $f_2(\alpha,\theta)$ represented by the infinite series (\ref{95}).
Fortunately, using the orthogonality relationship for the Legendre polynomials, we can get these integrals also in explicit form. Changing the order of summation and integration, we easily obtain:
\begin{eqnarray}\label{123}
I_{1,3}^{(2)}=\frac{\pi^{-3/2}}{3\sqrt{5}}\sum_{l=0}^\infty \int_0^\pi [\sin \alpha+3 \sin(3\alpha)]~\eta
\left[\frac{\zeta_l(\rho)}{(2l-1)(2l+3)}\right]\sin \alpha d\alpha
\int_0^\pi P_l(\cos \theta)\cos \theta \sin \theta d \theta
~\nonumber~\\
=\frac{2\pi^{-3/2}}{45\sqrt{5}}
\int_0^{\pi}[\sin \alpha+3 \sin(3\alpha)]~\eta~\zeta_1(\rho)\sin \alpha d \alpha=\frac{7\pi+22}{210\pi^{3/2}\sqrt{5}},~~~~~~~~~~~~~
\end{eqnarray}
\begin{eqnarray}\label{124}
I_{3,3}^{(2)}=\frac{4\pi^{-3/2}}{3\sqrt{5}}\sum_{l=0}^\infty \int_0^\pi \eta
\left[\frac{\zeta_l(\rho)}{(2l-1)(2l+3)}\right]\sin^4 \alpha~d\alpha
\int_0^\pi P_l(\cos \theta)P_3(\cos \theta) \sin \theta d \theta
~~~\nonumber~\\
=\frac{8\pi^{-3/2}}{945\sqrt{5}}
\int_0^{\pi}\eta~\zeta_3(\rho)\sin^4 \alpha~d \alpha=\frac{3\pi-32}{180\pi^{3/2}\sqrt{5}}.~~~~~~~~~~~~
\end{eqnarray}
Recall that $\eta\equiv \eta(\alpha)$ is defined by Eq.(\ref{6}) and $\rho=\tan(\alpha/2)$.

It can be shown (using fairly long non-trivial derivations) that the integrals $I_{l,2}$ and $I_{l,4}$ vanish both for $l=1$ and $l=3$. This means that (according to the representations (\ref{110}) and (\ref{111})) the AFC, $\psi_{6,3}(\alpha,\theta)$ is proportional to the third power of the nucleus charge $Z$ (only), which is in full agreement with formula (13) from Ref.\cite{LEZ4}.

Thus, combining the results of this Section, we obtain the nonzero coefficients $a_{6,l}$ in the simple final form:
\begin{equation}\label{125}
a_{61}=-\frac{(\pi-2)(5\pi-14)(32\pi-97)}{56700 \pi^{3/2}\sqrt{5}}Z^3,
\end{equation}
\begin{equation}\label{125}
a_{63}=-\frac{(\pi-2)(5\pi-14)(357\pi-1112)}{680400 \pi^{3/2}\sqrt{5}}Z^3.
\end{equation}

\section{Derivation of the angular Fock coefficients $\psi_{7,3}(\alpha,\theta)$ and $\psi_{8,4}(\alpha,\theta)$}\label{S3}

In Sections (\ref{S1}) and (\ref{S2}) we have detailed the derivation of the AFCs $\psi_{5,2}(\alpha,\theta)$ and $\psi_{6,3}(\alpha,\theta)$, respectively.
Therefore, for the AFCs $\psi_{7,3}(\alpha,\theta)$ and $\psi_{8,4}(\alpha,\theta)$ we give only abbreviated derivations, and include extended explanations only in case of significant differences.

\subsection{The AFC $\psi_{7,3}(\alpha,\theta)$}\label{S3a}

The FRR (\ref{3})-(\ref{4}) for $k=7$ and $p=3$ reduces to the form
\begin{equation}\label{131}
\left(\Lambda^2-77\right)\psi_{7,3}(\alpha,\theta)=h_{7,3}(\alpha,\theta),
\end{equation}
where
\begin{equation}\label{132}
h_{7,3}(\alpha,\theta)=-2V\psi_{6,3}(\alpha,\theta).
\end{equation}
Using Eqs.(\ref{110}), (\ref{110a}) and (\ref{5}) the RHS of Eq.(\ref{131}) can be represented in the form:
\begin{equation}\label{133}
h_{7,3}(\alpha,\theta)=\frac{(\pi-2)(5\pi-14) Z^3}{340200\sqrt{5}\pi^{3/2}}
\left\{\frac{\breve{h}_1+\breve{h}_2}{\sqrt{5}\pi^{3/2}}
-2Z\left[12(32\pi-97)\breve{h}_3+(357\pi-1112)\breve{h}_4\right]\right\},
\end{equation}
where
\begin{equation}\label{134}
\breve{h}_1=20 \xi^{-1}\left[12(32\pi-97)(1-\xi^2)+(357\pi-1112)(1-\xi^2)^3\right],
\end{equation}
\begin{equation}\label{135}
\breve{h}_2=60(688-255\pi) \xi^{-1}\sin^3\alpha\cos \theta,~~~~~~~~~~~~~~~~~~~~~~~~~~~~~~~~
\end{equation}
\begin{equation}\label{136}
\breve{h}_3=\eta \sin^{-1}\alpha Y_{6,1}(\alpha,\theta),~~~~~\breve{h}_4=\eta \sin^{-1}\alpha Y_{6,3}(\alpha,\theta).~~~~~~~~~~~~
\end{equation}
Accordingly, the solution of the FRR (\ref{131}) can be found in the form:
\begin{equation}\label{137}
\psi_{7,3}(\alpha,\theta)=\frac{(\pi-2)(5\pi-14) Z^3}{340200\sqrt{5}\pi^{3/2}}
\left\{\frac{\breve{f}_1+\breve{f}_2}{\sqrt{5}\pi^{3/2}}
-2Z\left[12(32\pi-97)\breve{f}_3+(357\pi-1112)\breve{f}_4\right]\right\},
\end{equation}
where the AFC-components $\breve{f}_i$ satisfy the IFRRs
\begin{equation}\label{138}
\left(\Lambda^2-77\right)\breve{f}_i=\breve{h}_i.~~~~~~~~~(i=1,2,3,4)
\end{equation}
Note that the components $\breve{h}_i$ of the RHS $h_{7,3}$ of the FRR (\ref{131}) for the AFC $\psi_{7,3}$
are reasonably close to the components $h_i$ of the RHS $h_{5,2}$ of the FRR (\ref{21}).
Therefore, we will only briefly dwell on the conclusions of the corresponding results, as we noted earlier.

It is seen from Eq.(\ref{134}) that the RHS $\breve{h}_1$ is a function of a single variable $\xi$ defined by Eq.(\ref{6}).
The solution of the corresponding IFRR have been described in Sec. IV of Ref. \cite{LEZ4} (see also Sec. II  of Ref. \cite{LEZ2}) and illustrated (among others) in Sec. \ref{S1c} of the current article. Thus, following the technique mentioned above, we obtain:
\begin{eqnarray}\label{139}
\breve{f}_1=\left(\frac{41437\pi}{12}-\frac{74342}{7}\right)\xi^7+\left(36476-\frac{35588\pi}{3}\right)\xi^5+
~~~~~~~~~~~~~~~~~~~\nonumber~\\
+\frac{5}{2}(4931\pi-15156)\xi^3+5(2276-741\pi)\xi.
\end{eqnarray}
It can be verified that the RHSs $\breve{h}_3$ and $\breve{h}_4$ represent functions of the form $f(\alpha)P_l(\cos \theta)$ with $l$ equals $1$ and $3$, respectively.
The solution of the corresponding IFRR have been described in Sec. V of Ref. \cite{LEZ4} and illustrated in Sections \ref{S1a} and \ref{S1b} of the current article. This enables us to obtain:
\begin{equation}\label{140}
\breve{f}_3=-\frac{\rho(1+\rho)\left(29+\rho\left\{16+\rho[\rho(16+29\rho)-114]\right\}\right)\cos \theta}{9\sqrt{5}~\pi^{3/2}(\rho^2+1)^{7/2}},
\end{equation}
\begin{equation}\label{141}
\breve{f}_4=-\frac{\sin^3\alpha \sqrt{1+\sin \alpha}}{2\sqrt{5}~\pi^{3/2}}P_3(\cos \theta).
\end{equation}
Remind that the $\rho$ variable was defined previously in Sec. \ref{S1a}.

The RHS $\breve{h}_2$ represented by Eq.(\ref{135}) is slightly more complicated than $h_2$ discussed in Sec. \ref{S1d}.
%Therefore we describe the solution of the IFRR (\ref{138}) with $i=2$ in more detail (but again without step-by-step derivations).
In this regard, it would be useful to clarify two points.

First, using representation (\ref{66}) for $\xi^{-1}$ we can rewrite Eq.(\ref{135}) in the form:
\begin{equation}\label{142}
\breve{h}_2=60(688-225\pi)\bar{h}_2
\end{equation}
where
\begin{equation}\label{143}
\bar{h}_2=\sum_{l=0}^\infty 2^{-l}(\sin \alpha)^{l+3}F_l(\rho)\cos \theta P_l(\cos \theta),~~~~~~~~~~~~~~~~~~~~~~~~~~~~~~~~
\end{equation}
and where $F_l(\rho)$ is defined by Eqs.(\ref{67})-(\ref{67a}).
%To reduce representation (\ref{142}) to the form which can be applied to solve
In order to apply to the solution of the corresponding IFRR by the method described in Sec. \ref{S1d} (see also \cite{LEZ4}), the $\theta$-dependent $l$-component in the series expansion of $\bar{h}_2$ must be pure $P_l(\cos \theta)$.
To solve the problem one could use the general formula representing the Clebsch-Gordan series for product of two \emph{spherical} harmonics.
However, in our simple case, it is easier to use the recurrence relation for the Legendre polynomials
\begin{equation}\label{144}
(l+1)P_{l+1}(x)-(2l+1)x P_l(x)+l P_{l-1}(x)=0,
\end{equation}
which enables us to represent $\bar{h}_2$ in the desired form:
\begin{equation}\label{145}
\bar{h}_2=\sum_{l=0}^\infty \bar{\textrm{h}}_l(\rho) (\sin \alpha)^l P_l(\cos \theta),~~~~~~~~~~~~~~~~~~~~~~~~~~~~~~~~
\end{equation}
where
\begin{equation}\label{146}
\bar{\textrm{h}}_l(\rho)=\frac{l}{2^{l-1}(2l-1)}\sin^2\alpha F_{l-1}(\rho)+\frac{l+1}{2^{l+1}(2l+3)}\sin^4\alpha F_{l+1}(\rho).
\end{equation}
The second point is related to the calculation of the coefficient
\begin{equation}\label{147}
\bar{h}_{2l,l}=\pi^2\int_0^{\pi}\int_0^{\pi}\bar{h}_2(\alpha,\theta)Y_{2l,l}(\alpha,\theta)\sin^2\alpha\sin \theta d\alpha d\theta
\end{equation}
in HH-expansion of $\bar{h}_2$ (see the corresponding Eq.(\ref{92}) for calculation of $\psi_{5,2}(\alpha,\theta)$).
Of course, we can use representation (\ref{145})-(\ref{146}) and then apply the orthogonality condition for the Legendre polynomials.
However, the simpler way is to use the original representation (\ref{143}) taking into account that $\cos \theta\equiv P_1(\cos \theta)$. In this case, we can apply the well-known formula for the integral of three Legendre polynomials
\begin{equation}\label{148}
\int_{-1}^1 P_l(x)P_L(x)P_{l'}(x)dx=2\begin{pmatrix}
l& L& l'\\
0& 0& 0 \\
\end{pmatrix}^2,
\end{equation}
where the RHS represents twice the square of the Wigner 3-j symbol.

Thus, applying the methodology outlined in Sec. \ref{S1d}, and given the above features, one obtains
\begin{equation}\label{149}
\bar{f}_2=\frac{1}{48}\sum_{l=0}^\infty \frac{\bar{\zeta}_l(\rho)P_l(\cos \theta)}{(2l-1)(2l+3)},
\end{equation}
where
\begin{equation}\label{150}
\bar{\zeta}_l(\rho)=
\left\{ \begin{array}{c}\mathlarger{\bar{\chi}_l(\rho) },~~~~~~~~~~0\leq\rho\leq1\\
\mathlarger{\bar{\chi}_l(1/\rho)}, ~~~~~~~~~~~~\rho\geq 1
\end{array}\right.
\end{equation}
with
\begin{eqnarray}\label{151}
\bar{\chi}_l(\rho)=-\frac{\rho^l}{(\rho^2+1)^{7/2}}\Big{\{}\frac{(32l^2+26l-25)\rho^6}{2l+5}\left[\frac{(2l-1)\rho^2}{2l+9}
+4\right]+
~~~~~~~~~~~~~~~~\nonumber~\\
+\frac{1}{2l-3}\left[
\frac{6(84l^2+84l-95)\rho^4}{2l+5}-(32l^2+38l-19)\left(\frac{2l+3}{2l-7}+4\rho^2\right)
\right]\Big{\}}.~~~~~~~~
\end{eqnarray}
Recall that the component $\breve{f}_2$ in the RHS of Eq.(\ref{137}) is equal to $60(688-225\pi)\bar{f}_2$ according to representation (\ref{142}).

As in the case of the AFC $\psi_{5,2}(\alpha,\theta)$, there are combinations of special hyperspherical angles $\alpha$ and $\theta$ for which the component $\bar{f}_2\equiv\bar{f}_2(\alpha,\theta)$ of the AFC $\psi_{7,3}(\alpha,\theta)$ can be obtained in closed form.
In particular, one obtains:
\begin{equation}\label{152}
\bar{f}_2(\alpha,0)=\mp \frac{(\rho-1)(95\rho^6+1166\rho^5-1879\rho^4-8844\rho^3-1879\rho^2+1166\rho+95)}{5040(\rho^2+1)^{7/2}},
\end{equation}
\begin{equation}\label{153}
\bar{f}_2(\alpha,\pi)=\frac{(\rho+1)(95\rho^6-1166\rho^5-1879\rho^4+8844\rho^3-1879\rho^2-1166\rho+95)}{5040(\rho^2+1)^{7/2}},
\end{equation}
\begin{equation}\label{154}
\bar{f}_2(\alpha,\frac{\pi}{2})=\frac{19\rho^4+10\rho^2+19}{1008(\rho^2+1)^2}.
\end{equation}
Sign "$-$" in Eq.(\ref{152}) corresponds to $0\leq\alpha\leq\pi/2$ ($0\leq \rho\leq 1$), whereas "$+$" to $\pi/2\leq\alpha\leq\pi$ ($ \rho\geq 1$).

For the important cases of the nucleus-electron and electron-electron coalescence, representation (\ref{149})-(\ref{151}) is simplified to:
\begin{equation}\label{155}
\bar{f}_2\left(0,\theta\right)=\frac{19}{1008},~~~~~~~~~~~~~\bar{f}_2\left(\frac{\pi}{2},0\right)=0.
\end{equation}

\subsection{The AFC $\psi_{8,4}(\alpha,\theta)$}\label{S3b}
%xxxxxxxxxxxxxxxxxxxxxxxxxxxxxxxxxxxxxxxxxxxxxxxxxxxxxxxxx

Having at our disposal the AFC $\psi_{7,3}\equiv\psi_{7,3}(\alpha,\theta)$, we can calculate the AFC $\psi_{8,4}\equiv\psi_{8,4}(\alpha,\theta)$ using the FRR
(\ref{3})-(\ref{4}) for $k=8$ and $p=3$:
\begin{equation}\label{161}
\left(\Lambda^2-96\right)\psi_{8,3}=80 \psi_{8,4}-2 V \psi_{7,3}+2E \psi_{6,3}.
\end{equation}
It follows from Eq.(\ref{89a}) and the FRR (\ref{3})-(\ref{4}) for $k=8$ and $p=4$ that the AFC, $\psi_{8,4}$ is a linear combination of the HHs, $Y_{8,l}\equiv Y_{8,l}(\alpha,\theta)$.
Moreover, given that only $Y_{n,l}(\alpha,\theta)$ with even values of $n/2-l$ are suitable for singlet S-states, we obtain:
\begin{equation}\label{162}
\psi_{8,4}=a_{80}Y_{8,0}+a_{82}Y_{8,2}+a_{84}Y_{8,4}.
\end{equation}
For further derivations, it is advisable to represent the HHs in the form
\begin{equation}\label{163}
Y_{8,l}(\alpha,\theta)=y_{8l}(\alpha)P_l(\cos \theta),
\end{equation}
where
\begin{equation}\label{164}
y_{80}(\alpha)=\pi^{-3/2}\left[2\cos(4\alpha)+2\cos(2\alpha)+1\right],
\end{equation}
\begin{equation}\label{165}
y_{82}(\alpha)=\frac{2}{\pi^{3/2}}\sqrt{\frac{10}{7}}\sin^2\alpha\left[4\cos(2\alpha)+3\right],
\end{equation}
\begin{equation}\label{166}
y_{84}(\alpha)=\frac{8}{\pi^{3/2}}\sqrt{\frac{2}{7}}\sin^4\alpha.
\end{equation}
It was found in Sec. \ref{S2} that $\psi_{6,3}\equiv\psi_{6,3}(\alpha,\theta)$ is the linear combination of the HHs $Y_{6,l}(\alpha,\theta)$.
Thus, expanding each function of Eq.(\ref{161}) in HHs, and equating the coefficients for $Y_{8,l}$, we obtain
(see the corresponding result (\ref{109}) for $a_{6l}$)
\begin{equation}\label{167}
a_{8l}=\frac{1}{40}\int V \psi_{7,3}Y_{8,l} d \Omega,
\end{equation}
where the potential $V$ is defined by Eq.(\ref{5}).
When deriving the last equation, it was taken into account that $(\Delta^2-96)Y_{8,l}=0$, as follows from Eq.(\ref{89a}).

Direct substitution of the representations (\ref{5}), (\ref{137}) and (\ref{163}) into the RHS of Eq. (\ref{167}), yields:
\begin{eqnarray}\label{168}
a_{8l}=\frac{(\pi-2)(5\pi-14)Z^3}{13608000\sqrt{5}~\pi^{3/2}}\times
~~~~~~~~~~~~~~~~~~~~~~~~~~~~~~~~~\nonumber~~~~~~~~~~~~~~\\
\times\int\left(\frac{1}{\xi}-\frac{2Z\eta}{\sin \alpha}\right)
\left\{\frac{\breve{f}_1+\breve{f}_2}{\sqrt{5}\pi^{3/2}}
-2Z\left[12(32\pi-97)\breve{f}_3+(357\pi-1112)\breve{f}_4\right]\right\}Y_{8,l}(\alpha,\theta)d\Omega
.~\nonumber~\\
\end{eqnarray}
It follows from Eq.(13) of Ref.\cite{LEZ4} that only the coefficients at $Z^4$ are nonzero on the RHS of the last equation. Whence, Eq.(\ref{168}) reduces to the form:
\begin{equation}\label{169}
a_{8l}=-\frac{(\pi-2)(5\pi-14)Z^4}{6804000\sqrt{5}~\pi^{3/2}}
\left[\frac{S_{1l}+S_{2l}}{\sqrt{5}\pi^{3/2}}
+12(32\pi-97)S_{3l}+(357\pi-1112)S_{4l}\right],
\end{equation}
where
\begin{equation}\label{170}
S_{1l}=\pi^2\int_0^\pi\int_0^\pi \breve{f}_1(\xi)Y_{8,l}(\alpha,\theta)\eta \sin \alpha \sin \theta d\alpha d\theta,
\end{equation}
\begin{eqnarray}\label{171}
S_{2l}=\pi^2\int_0^\pi\int_0^\pi \breve{f}_2(\alpha,\theta)Y_{8,l}(\alpha,\theta)\eta \sin \alpha \sin \theta d\alpha d\theta=
~~~~~~~~~~~~~~~~~~~~~~~~~~~~~~~~~\nonumber~~~~~~~~~~~~~~\\
=\frac{5\pi^2(688-225\pi)}{(2l-1)(2l+1)(2l+3)}\int_0^{\pi/2}\bar{\chi}_l(\rho)y_{8l}(\alpha)\eta\sin \alpha d\alpha,~~~~~~
\end{eqnarray}
\begin{equation}\label{172}
S_{nl}=\pi^2\int_0^\pi\int_0^\pi \breve{f}_n(\alpha,\theta)Y_{8,l}(\alpha,\theta)\xi^{-1} \sin^2 \alpha \sin \theta d\alpha d\theta.~~~~~~~~~~(n=3,4)
\end{equation}
The identifiers $\xi$ and $\eta$ are defined by Eq.(\ref{6}), whereas functions $\bar{\chi}_l(\rho)$ can be calculated by the formula (\ref{151}). When deriving Eq.(\ref{171}) we applied the orthogonality condition for the Legendre polynomials. Fortunately, all integrals (\ref{170})-(\ref{172}) can be taken in closed form.
Thus, by collecting these results and substituting them into the RHS of Eq.(\ref{169}), we finally obtain the desired coefficients in the form:
\begin{equation}\label{173}
a_{8l}=\frac{Z^4(\pi-2)(5\pi-14)}{\pi^{5/2}}b_{8l},
\end{equation}
with
\begin{eqnarray}\label{174}
b_{80}=\frac{\pi(150339\pi-927292)+1430792}{19289340000},~~~
b_{82}=\frac{\pi(751965\pi-4654046)+7200976}{1928934000\sqrt{70}},
~~~~\nonumber~~~~~~~\\
b_{84}=\frac{\pi(3190317\pi-19828996)+30802176}{25719120000\sqrt{14}}.~~~~~~~~~~~~~~~~~~~~~~~~~
\end{eqnarray}

\section{Results and Discussions}\label{S4}

The angular Fock coefficients $\psi_{k,p}\equiv\psi_{k,p}(\alpha,\theta)$ with the \emph{maximum possible} value of subscript $p$ were calculated on examples of the coefficients with $5\leq k \leq 8$.
The AFCs $\psi_{9,4}$ and $\psi_{10,5}$ are additionally presented in the Appendix without derivations.
The presented technique makes it possible to calculate such AFCs for any arbitrarily large $k$.
These coefficients are leading in the logarithmic power series representing the Fock expansion (see Eq.(\ref{8})). As such, they may be indispensable for the development of simple methods for calculating the helium-like electronic structure.

The proposed technique, as well as the final results, are quite complex. Therefore, both require verification.
We are aware of two ways of the above-mentioned verification.
The first one is to use the Green's function (GF) approach (see, Ref. \cite{FOCK} and also Ref. \cite{LEZ7}, Sec. 4) which enables us to calculate (at least, numerically) the AFC (or its component) by the following integral representation:
\begin{equation}\label{175}
\psi_{k,p}(\alpha,\theta)=\frac{1}{8\pi}\int_0^\pi d\alpha' \sin^2\alpha'\int_0^\pi d\theta' \sin \theta'~ h_{k,p}(\alpha',\theta')
\int_0^\pi \frac{\cos\left[\left(\frac{k}{2}+1\right)\omega \right]}{\sin \omega}(1-\lambda)d\varphi,
\end{equation}
where $\omega$ is an angle defined by the relation
\begin{equation}\label{176}
\cos \omega =\cos \alpha \cos \alpha'+\sin \alpha \sin \alpha'
\left(\cos \theta \cos \theta'+\sin \theta \sin \theta' \cos \varphi \right),
\end{equation}
%with auxiliary angle $\varphi \in [0,2\pi]$.
whereas
\begin{equation}\label{177}
\lambda=
\left\{ \begin{array}{c}
0~~~~~~~~~~~~~~k~~~odd\\
\omega/\pi~~~~~~~~~~~~k~~~even\\
\end{array}\right..
\end{equation}
For even $k$ and maximum value of $p=k/2$ the RHS $h_{k,k/2}$ of the FRR (\ref{3}) equals zero.
This implies that the GF formula (\ref{175}) cannot be applied in this case.
Hence, only the AFCs $\psi_{k,p}$ with odd values of $k$ (and maximum $p$) can be verified with the GF method.
Thus, numerically calculating (for various combinations of angles $\alpha$ and $\theta$) the triple integrals (\ref{175}) representing the AFCs $\psi_{5,2}(\alpha,\theta)$, $\psi_{7,3}(\alpha,\theta)$ and $\psi_{9,4}(\alpha,\theta)$,
we have verified that the representations obtained for them in Sections \ref{S1},  \ref{S3a} and in the Appendix are correct.
%Note that the GF approach admits the calculation of not only the AFCs but also the AFC components representing the factors for a definite power of the nucleus charge $Z$.

The second method of the verification under consideration, covering all possible combinations of angles, is much more complicated. This is the CFHH method mentioned in Introduction.
It is based on decomposing the full WF into a form
\begin{equation}\label{180}
\Psi^{\textrm{CFHH}}(r_1,r_2,r_{12})=\exp\left[f(r_1,r_2,r_{12})\right]\Phi^{\textrm{CFHH}}(R,\alpha,\theta),
\end{equation}
where the so called correlation function $f$ can be taken in a simple linear form
\begin{equation}\label{181}
f(r_1,r_2,r_{12})=c_1 r_1+c_2 r_2+c_{12}r_{12}.
\end{equation}
The so called "cusp parametrization"
\begin{equation}\label{182}
c_1=c_2=-Z,~~~~~~~c_{12}=1/2
\end{equation}
is used as a rule.
For small enough hyperspherical radius $R$, the function $\Phi$ is represented as
\begin{equation}\label{183}
\Phi^{\textrm{CFHH}}(R,\alpha,\theta)=\frac{1}{d_{0,0}(\alpha,\theta)}\sum_{k=0}^K (2\kappa R)^k
\sum_{p=0}^{[k/2]}d_{k,p}(\alpha,\theta)\ln^p(2 \kappa R),
\end{equation}
where $\kappa=\sqrt{-2E}$, and functions $d_{k,p}(\alpha,\theta)$ are expanded in $N$ (basis size) HHs. It follows from representation (\ref{183}) that the AFCs $\psi_{k,p}(\alpha,\theta)$ can be expressed in terms of the functions $d_{k',p'}(\alpha,\theta)$ calculated by the CFHHM.
For example, for the AFCs in questions one obtains:
\begin{equation}\label{184}
\psi_{k,k/2}^{\textrm{CFHH}}(\alpha,\theta)=\frac{(2\kappa)^k d_{k,k/2}(\alpha,\theta)}{d_{0,0}(\alpha,\theta)}.
\end{equation}
We have calculated all AFCs discussed in this article using CFHHM with $K=18$ and $N=1600$.
The angles $0\leq\alpha\leq\pi$ and $0\leq\theta\leq\pi$ with step $\pi/6$ were considered.
The relative difference $|1-\psi_{k,p}(\alpha,\theta)/\psi_{k,p}^{\textrm{CFHH}}(\alpha,\theta)|$ was less than $10^{-7}$ for all examined cases, including $1\leq Z\leq 5$. This indicates that all our theoretical calculations were correct.

%\section{Acknowledgments}\label{S5}
%This work was supported by the PAZY Foundation, Israel.
\appendix

\section{}\label{SA}

In Sections \ref{S1}-\ref{S2} the AFCs $\psi_{5,2}(\alpha,\theta)$ and $\psi_{6,3}(\alpha,\theta)$ were calculated with detailed derivations.
In Sec. \ref{S3} the AFCs $\psi_{7,3}(\alpha,\theta)$ and $\psi_{8,4}(\alpha,\theta)$ were presented with a very brief derivations.
In this Appendix, we present the AFCs $\psi_{9,4}\equiv\psi_{9,4}(\alpha,\theta)$ and  $\psi_{10,5}\equiv\psi_{10,5}(\alpha,\theta)$ without derivation. To find the latter AFCs, the methods described in the main sections were used.

So, the first AFC under consideration can be represented as:
\begin{equation}\label{A1}
\psi_{9,4}=2Z^4\left[2Z X_1(\alpha,\theta)-X_2(\alpha,\theta)\right],
\end{equation}
where
\begin{equation}\label{A2}
X_1(\alpha,\theta)=\check{a}_{80}\check{f}_1+\check{a}_{82}\check{f}_2+\check{a}_{84}\check{f}_3,
\end{equation}
\begin{equation}\label{A3}
X_2(\alpha,\theta)=\frac{35}{\pi^{3/2}}\sqrt{\frac{2}{7}}~\check{a}_{84}\check{f}_4+
\frac{(\pi-2)(5\pi-14)}{123451776000 \pi^4}
\left[c_{5}\check{f}_5+c_{6}\check{f}_6+c_{7}\check{f}_7+16(c_{8}\check{f}_8+c_{9}\check{f}_9)\right].
\end{equation}
Here $\check{a}_{8l}=Z^{-4}a_{8l}$, where the coefficients $a_{8l}$ are defined by Eqs.(\ref{173}), (\ref{174}), whereas the other coefficients are:
\begin{eqnarray}\label{A4}
c_{5}=\pi(29757524-4780401 \pi)-46286848,~~~
c_{6}=\pi(9581100\pi-59458928)+92239360,~~~~
\nonumber~\\
c_{7}=\pi(28060+10149 \pi)-167168,~~~~
c_{8}=9\pi(134543 \pi-828732)+11488128,~~~~~~~
\nonumber~\\
c_{9}=\pi(4804833 \pi-29773780)+46119680.~~~~~~~~~~~~~~~~~~~~~~~~~~~~~~~~~~~~~~~~~~~~~
\end{eqnarray}
The functions $\check{f}_i\equiv\check{f}_i(\alpha,\theta)$ are:
\begin{equation}\label{A5}
\check{f}_1=-\frac{(\rho+1)(563 \rho^8+1012 \rho^7-8932 \rho^6-3668 \rho^5+23954 \rho^4-3668 \rho^3-8932 \rho^2+1012 \rho+563)}{1260 \pi^{3/2} (\rho^2+1)^{9/2}},
\end{equation}
\begin{equation}\label{A6}
\check{f}_2=-\left(\frac{8}{\pi^{3/2}}\sqrt{\frac{10}{7}}\right)
\frac{\rho^2 (1+\rho) (126+49 \rho-424 \rho^2+49 \rho^3+126 \rho^4)}{300 (\rho^2+1)^{9/2}}P_2(\cos \theta),
\end{equation}
\begin{equation}\label{A7}
\check{f}_3=-\frac{2}{5\pi^{3/2}}\sqrt{\frac{2(1+\sin \alpha)}{7}}
\sin^4 \alpha~P_2(\cos \theta),
\end{equation}
\begin{equation}\label{A8}
\check{f}_4=-\frac{\xi(315-1680 \xi^2+2814 \xi^4-1854 \xi^6+419 \xi^8)}{1260},
\end{equation}
\begin{equation}\label{A9}
\check{f}_5=-\frac{\xi}{60}(2 \xi^2-3) (2 \xi^2-1) (4 \xi^4-10 \xi^2+5).
\end{equation}
The remaining $\check{f}$-functions are represented by series:
\begin{equation}\label{A10}
\check{f}_j=\frac{(\rho^2+1)^{-9/2}}{k_j}\sum_{l=0}^\infty \frac{\rho^l\zeta_{jl}(\rho)}{(2l-1)(2l+3)}P_l(\cos \theta),
~~~~~~~~~~~~(j=6,7,8,9)
\end{equation}
where
\begin{equation}\label{A11}
k_6=6,~~~~~~k_7=60,~~~~~~k_8=24,~~~~~~k_9=40,
\end{equation}
and the corresponding $\zeta$-functions are:
\begin{eqnarray}\label{A12}
\zeta_{6l}(\rho)=\frac{(2l-15) (2l-1) (l+1) \rho^{10}}{(2l+7) (2l+11)}+
\frac{(22l^2-5l-12)\rho^8}{(2l+7)}+\frac{10(2l^2+11l+3)\rho^6}{2l+7}-
\nonumber~\\
-\frac{10 (2l^2-7l-6)\rho^4}{2l-5}-\frac{(22l^2+49l+15)\rho^2}{2l-5}-\frac{l(2l+3)(2l+17)}{(2l-9)(2l-5)}
,~~~~~~~~~~~
\end{eqnarray}
\begin{eqnarray}\label{A13}
\zeta_{7l}(\rho)=\frac{(2l-1)(4l^2+160l-189) \rho^{10}}{(2l+7) (2l+11)}+
\frac{35(4l^2+40l-9)\rho^8}{(2l+7)}-\frac{350(4l^2+16l+3)\rho^6}{2l+7}+
\nonumber~\\
+\frac{350(4l^2-8l-9)\rho^4}{2l-5}-\frac{35(4l^2-32l-45)\rho^2}{2l-5}-\frac{(2l+3)(4l^2-152 l-345)}{(2l-9)(2l-5)}
,~~~~~~~~~~~
\end{eqnarray}
\begin{eqnarray}\label{A14}
\zeta_{8l}(\rho)=-\frac{(2l-1)(56l^3+250l^2+338l+171)\rho^{10}}{(2l+5)(2l+7) (2l+11)}-
\frac{(136l^3+314l^2-110l-153)\rho^8}{(2l+5)(2l+7)}-
~\nonumber~~~~~~\\
-\frac{2(80l^4+652l^3+566l^2-1824l-873)\rho^6}{(2l-3)(2l+5)(2l+7)}
+\frac{2(80l^4-332l^3-910l^2+1320l+945)\rho^4}{(2l-5)(2l-3)(2l+5)}+
~\nonumber~~\\
+\frac{(136l^3+94l^2-330l-135)\rho^2}{(2l-5)(2l-3)}+\frac{(2l+3)(56l^3-82l^2+6l-27)}{(2l-9)(2l-5)(2l-3)}
,~~~~~~~~~~~~~~~~
\end{eqnarray}
\begin{eqnarray}\label{A15}
\zeta_{9l}(\rho)=\frac{(2l-1)(24l^3-150l^2-670l-439)\rho^{10}}{(2l+5)(2l+7) (2l+11)}+
\frac{5(72l^3+162l^2-70l-103)\rho^8}{(2l+5)(2l+7)}+
~\nonumber~~~~~~\\
+\frac{10(16l^4+220l^3+222l^2-804l-423) \rho^6}{(2l-3)(2l+5)(2l+7)}
-\frac{10(16l^4-156l^3-342l^2+652l+399)\rho^4}{(2l-5)(2l-3)(2l+5)}-
~\nonumber~~\\
-\frac{5(72l^3+54l^2-178l-57)\rho^2}{(2l-5)(2l-3)}-\frac{(2l+3)(24l^3+222l^2-298l-57)}{(2l-9)(2l-5)(2l-3)}
,~~~~~~~~~~~~~~~~
\end{eqnarray}
It is important to emphasize that the representations (\ref{A10})-(\ref{A15}) are valid only for $0\leq \rho \leq1$.
For values $\rho>1$, one should replace $\rho$ with $1/\rho$, which is equivalent to simply redefining $\rho$ as $\cot(\alpha/2)$.

The second AFC under consideration is of the form:
\begin{equation}\label{A20}
\psi_{10,5}=-\frac{Z^5(\pi-2)(5\pi-14)}{\pi^{7/2}}
\left[b_{10,1}Y_{10,1}(\alpha,\theta)+b_{10,3}Y_{10,3}(\alpha,\theta)+b_{10,5}Y_{10,5}(\alpha,\theta)\right],
\end{equation}
where
\begin{equation}\label{A21}
b_{10,1}=\frac{ \pi[3 \pi (6840010557 \pi-63828704998)+595609133656]-617517605744}
{401025378600000 \sqrt{105}},
\end{equation}
\begin{equation}\label{A23}
b_{10,3}=\frac{ \pi [\pi (9194460432 \pi-85833963053)+267084629592]-277009842768}
{100256344650000 \sqrt{30}},
\end{equation}
\begin{equation}\label{A25}
b_{10,5}=\frac{\pi [\pi(622341848670 \pi-5812646794643)+18095537797140]-18776793358080}
{10025634465000000 \sqrt{42}},
\end{equation}
and $Y_{10,l}(\alpha,\theta)$ with $l=1,3,5$ are the normalized HHs.

\newpage
%\renewcommand{\arraystretch}{0.85}

%\begin{figure}
%\centering
%\caption{The coefficient $a_{21}(Z)$ of the Fock expansion for the ground state of the helium atom and the two-electron ions %with infinitely massive nucleus of the charge $Z$.}
%\includegraphics[width=7.0in]{Fig_1.pdf}
%\label{F1}
%\end{figure}

%\bibliography{mybib}
\end{document}